\documentclass[article]{elsarticle}

\usepackage[russian, english]{babel}
\usepackage{graphicx}
\usepackage{amsmath}
\usepackage{amsthm}
\usepackage{amsfonts}
\usepackage{amssymb}
\usepackage{mathptmx}
\usepackage{eucal}
\usepackage{bm}

\newcommand*{\mailto}[1]{\href{mailto:#1}{\nolinkurl{#1}}}

\newtheorem {Lemma}{Lemma}[section]

\renewcommand{\Im}{\mathop{\mathrm{Im}}\nolimits}

\newcommand{\tr}{\mathop{\mathrm{tr}}\nolimits}

\numberwithin{equation}{section}

\begin{document}

\renewcommand{\figurename}{Fig.}
\renewcommand\bibname{References}

\title{Nonlinear dynamics of solitons for the vector modified Korteweg-de Vries equation}

\author[ilt]{Volodymyr Fenchenko\corref{cor1}} \ead{vfenchenko@gmail.com}
\author[ilt]{Evgenii Khruslov}
\cortext[cor1]{Corresponding author}
\address[ilt]{B.Verkin Institute for Low Temperature Physics and Engineering,\\* 61103
Kharkov, National Academy of Sciences, Ukraine}

\begin{abstract}
We consider the vector generalization of the modified Korteweg-de Vries equation. We  develop the inverse scattering transform  for solving this equation. We construct the solitons and the breather solutions and investigate the processes of their collisions. We show that along with one-component soliton solutions, there  are solutions which have essentially three-component structure.\end{abstract}

\begin{keyword}
vector mKdV, inverse scattering transform, soliton, collision, interaction profile.

\PACS 05.45.Yv
\end{keyword}

\maketitle

\section{Introduction.}

The Korteweg-de Vries  equation (KdV)
\begin{equation}
  \partial_t u - 6 u \partial_x u - \partial_x^3 u = 0
\end{equation}
is the classical equation of the theory of the nonlinear waves.
It was integrated for the first time in \cite{1} by the inverse scattering transform method.
Later the same method was applied to integrate other physically interesting equations.
Their soliton and multi-soliton solutions were found and interactions of these solutions were studied (see., e.g, \cite{2,3,4,7}). 	

Among these equations there is the  modified Korteweg-de Vries (mKdV) equation
\begin{equation}\label{1}
  \partial_t u + \alpha u^2 \partial_x u - \partial_x^3 u = 0
\end{equation}
It describes the nonlinear waves in media with dispersion, and it is completely integrable.
The dinamics of solutions for this equation depends essentially of the  sign of nonlinearity $\alpha$, $\pm \alpha>0. $
In particular, the soliton solutions exist for $\alpha>0$ only.
	
Along with the scalar mKdV equation (\ref{1}), the vector generalizations of this equation
\begin{equation*}
  \partial_t \mathbf{u} + 24 |\mathbf{u}|^2 \partial_x \mathbf{u} + \partial^3_x \mathbf{u} = 0, \;
  \partial_t \mathbf{u} + 12 \partial_x (|\mathbf{u}| \mathbf{u}) |\mathbf{u} |+ \partial^3_x \mathbf{u} = 0, \quad \mathbf u\in\mathbb R^n,\ n\geq2,
\end{equation*}
are also explored (see., e.g, \cite{3,9,10}) .
One has to mention that not all of these generalizations  are completely integrable.

In this paper we consider another vector generalization of the mKdV equation (\ref{1}) with  nonlinearity $\alpha=6$:
\begin{equation}\label{2}
  \partial_t \mathbf{u} + 6 \mathbf{u}\times \partial_x^2 \mathbf{u} + 6(\mathbf{u} \partial_x |\mathbf{u}|^2 - |\mathbf{u}|^2 \partial_x\mathbf{u})  + \partial^3_x \mathbf{u} = 0, \quad \mathbf u\in \mathbb R^3,
\end{equation}
where  $\mathbf{u}(x,t) $  is a three-component vector function of variables  $x\in\mathbb R$, $t\in\mathbb R$, and the sign  "$\times$" \ means the vector product in $\mathbb R^3$. It turns out that this equation is  completely integrable.	
	
More general equations of this type
\begin{equation*}
  \partial_t \mathbf{u} + a \mathbf{u}\times \partial_x^2 \mathbf{u} + b(\mathbf{u} \partial_x |\mathbf{u}|^2 - |\mathbf{u}|^2 \partial_x\mathbf{u})  + c \partial^3_x \mathbf{u} = 0,
\end{equation*}
with constant coefficients  $ a, b, c$ satisfying  $a^2 = 6 b c$, \; ($ b>0 $) a simple change of variables can be reduced to the canonical form (\ref{2}).The first two summands of these equations are the same  as in the Ginzburg-Landau equation.
They  appear in modelling of the nonlinear magnetization waves in ferromagnetic media \cite{5}.	

The aim of this paper is to develop the inverse scattering transform for equation (\ref{2}), to construct  the explicit formulas for some soliton solutions and to study their interactions.

\section{Reduction of the equation (\ref{2}) to the matrix equation. }
In \cite{6} it was mentioned, that equation (\ref{2}) can be reduced to the matrix completely integrable equation.
Namely, let
\begin{equation}\mathbf{u}(x,t) = (u_1(x,t), u_2(x,t), u_3(x,t)) \label{vec}\end{equation}  be a real-valued solution of equation (\ref{2}).  Put
\begin{equation}\label{3}
  \Phi(x,t) = -\mathrm i( u_1(x,t)\sigma_1 +  u_2(x,t)\sigma_2 +  u_3(x,t)\sigma_3),
\end{equation}
where
  $\sigma_1 = \begin{pmatrix} 0 &1 \\ 1 & 0 \end{pmatrix}$,
  $\sigma_2 = \begin{pmatrix} 0 &-\mathrm i \\ \mathrm i & 0 \end{pmatrix}$,
  $\sigma_3 = \begin{pmatrix} 1 & 0 \\ 0 & -1 \end{pmatrix}$
are the Pauli matrices and  $\mathrm i = \sqrt{-1}$. By straightforward  computation one can check that matrix-function (\ref{2}) satisfies the differential equation
\begin{equation}\label{4}
  \partial_t \Phi + 3[\Phi , \partial_x^2 \Phi] - 6 \Phi (\partial_x \Phi) \Phi - \partial_x^3 \Phi = 0,
\end{equation}
where $[\; ,\; ]$ \;   is the commutator.
On the other hand, if the complex valued matrix solution of equation (\ref{4}) satisfies conditions
\begin{equation}\label{5}
  \Phi^* = -\Phi, \quad \tr\Phi = 0,
\end{equation}
then this solution  can be represented as in (\ref{3}) with real functions
$u_1(x,t)$, $u_2(x,t)$, $u_3(x,t)$. Moreover, the vector  \eqref{vec} with these functions as components
satisfies equation (\ref{2}).
In (\ref{5}) the star denotes the complex conjugation and transpose, and "$\tr$" \  is the trace of the matrix.

Conditions (\ref{5}) are invariant under the action of the semigroup generated by equation (\ref{4}).
Namely, the following Lemma holds.

\begin{Lemma}\label{lem1}
Let    $\Phi(x,t)$  be a solution of the Cauchy problem  (\ref{5}) with the initial condition $\Phi(x,0)=\Phi_0(x)$ which satisfies condition (\ref{5}). Then $\Phi(x,t)$  also satisfies conditions (\ref{5}).
\end{Lemma}
The proof of this Lemma is straightforward and we omit it.

In \cite {6} the matrix equation (\ref{4}) is investigated in the class of real-valued solutions $\Phi(x,t)$.
By use of the inverse scattering transform it is proved there that equation (\ref{4}) is completely integrable.
With the same approach one can show that this results are valid for complex valued $\Phi(x,t)$.
In the next section we list without proof those of the results of them \cite {6} which are necessary for our further investigations.

\section{The inverse scattering transform method.}
Matrix equation (\ref{4}) admits the Lax representation
\begin{equation}\label{6}
  \partial_t \mathcal L = [\mathcal L,\mathcal A]
\end{equation}
where  $[\; , \;]$ is the commutator, and $L$ and $A$ are matrix differential operators
\begin{align*}
  & \mathcal L = \mathbb I \partial_x^2 + \Phi(x,t) \partial_x, \\
  & \mathcal A = 4 \mathbb I\partial_x^3 + 12 \Phi(x,t) \partial_x^2 + 6(\partial_x \Phi(x,t) +\Phi^2(x,t))\partial_x,
\end{align*}
and $\mathbb I$  is the identity  $3\times3$  matrix.
	
Consider the spectral equation
\begin{equation}\label{7}
  (\mathcal L+k^2 \mathbb I) U \equiv \partial_x^2 U + 2 \Phi(x,t) \partial_x U + k^2 U = 0,\; x\in\mathbb R,
\end{equation}
where  $t$  is fixed.
In what follows we assume that $\Phi(x,t)$   tends to $0$  sufficiently fast as  $x\rightarrow\pm\infty$  such that
\begin{equation*}
  \int^\infty_{-\infty}\left ( (1+x^2)|\Phi(x,t)|+(1+|x|)|\frac{\partial}{\partial x}\Phi(x,t)| \right) d x <\infty,\; \forall t>0.
\end{equation*}

In such a case there exists a matrix solution of  equation (\ref{6}) which can be represented as(\ref{6})
\begin{equation}\label{8}
U_{\pm}(x,k,t)=e^{\pm i k x} \mathbb I \pm i k \int_x^{\pm \infty} A^{\pm}(x,y,t) e^{\pm i k y} d y,
\end{equation}
where matrices  $A^{+}(x,y,t)$ and $A^{-}(x,y)$  belong to the spaces  $L_1(x,\infty)$ and  $L_1(-\infty,x,t)$   with respect to $y$.
The  matrix  $\Phi(x,t)$  can be computed  as (recall that $t$ is a parameter )
\begin{equation}\label{9}
  \Phi(x,t) = \pm \cfrac{d A^{\pm}(x,x,t)}{dx} \left(\mathbb I \mp A^{\pm}(x,x,t)\right)^{-1}
\end{equation}

For $\Im k = 0$  the matrix-functions $U_+(x,k,t)$ and $U_+(x,-k,t)$  form a fundamental system of solutions of (\ref{7}).
Then
\begin{equation}\label{10}
  U_-(x,k,t)=U_+(x,k,t) C_{11}(k,t)+U_+(x,-k,t) C_{12}(k,t),
\end{equation}
where the matrices  $C_{11}(k,t), C_{12}(k,t)$  are independent of $x$.
Note that the matrix  $C_{11}(k,t)$ is continuous on $\mathbb R$ and $C_{11}(k,t)=O(|k|^{-1})$  as $|k|\rightarrow\infty$.
The matrix  $C_{12}(k,t)$ is continuous on $\mathbb R$ and can be continued analytically into the upper half-plane $\mathbb C^+$ with $C_{12}(k,t)=C(t)+O(|k|^{-1})$   as $|k|\rightarrow\infty, \; \Im k\geq0$.

We assume  that  $\det C_{12} \neq 0$  for real  $k$. Consider matrices
\begin{align*}
  &S_{12}(k,t) = C_{11}(k,t)C_{12}^{-1}(k,t), \;  &\ \mbox{as}\ \ \; \Im \,k=0,\\
  &S_{11}(k,t) = C_{12}^{-1}(k,t), \;  &\ \mbox{as }\ \ \;\Im \,k\geq0,
\end{align*}
which are the reflection and transmission matrices respectively according to (\ref{8}) and (\ref{10}).

The matrix   $S_{11}(k,t)$ is meromorphic in $\mathbb C^+$, its  poles do not depend on $t$ and are located at the points  $k_\nu, \; (\Im k_\nu>0), \; \nu=1,2,...,N$ such that  $\lambda_\nu=k_\nu^2$  are eigenvalues of the the operator  $\mathcal L$.

The poles $k_\nu$  can be multiple and in their neighborhoods  $G_\nu\subset \mathbb C^+$  the decomposition  is valid
\begin{equation*}
  S_{11}(k,t)=(k-k_\nu)^{-n_\nu}S_{n_\nu}^{\nu}(t) +...+(k-k_\nu)^{-1}S_{1}^{\nu}(t) + S_0^{\nu}(k,t).
\end{equation*}
Here $n_\nu$ is the order of the pole $k_\nu$, $S_l^{\nu}(t), \; l=1,2,...,n_\nu$  are the constant matrices with respect to $k$,  $S_0^{\nu}(k,t)$  is a holomorphic matrix-function in the domain $G_\nu$.
One can show that there exist  matrices $R_1^{\nu}(t),...,R_{n_\nu}^{\nu}(t),$ $(R_{n_\nu}^{\nu} (t)\neq0)$ ,  which satisfy equalities (we omit  the dependence on $t$ in \eqref{11})
\begin{align}\label{11}
  &U_{-}(x,k_\nu) S_{n_\nu}^{\nu} = U_{+}(x,k_\nu) R_{n_\nu}^{\nu},\notag \\
  &U_{-}(x,k_\nu) S_{n_\nu - 1}^{\nu} +U^{(1)}_-(x,k_\nu) S_{n_\nu}^{\nu} = U_{+}(x,k_\nu) R_{n_\nu - 1}^{\nu} +U^{(1)}_-(x,k_\nu) R_{n_\nu}^{\nu},\notag \\
  &.......................................................................................... \\
  &U_{-}(x,k_\nu) S_1^{\nu} + U_-^{(1)}(x,k_\nu) S_2^{\nu} +...+\frac{1}{(n_\nu-1)!}
U^{ (n_\nu - 1)}_-(x,k_\nu) S_{n_\nu}^{\nu} = \notag \\ &= U_{+}(x,k_\nu) R_1^{\nu} + U^{(1)}(x,k_\nu) R_2^{\nu} +...+\frac{1}{(n_\nu-1)!}U^{(n_\nu - 1)}_+(x,k_\nu) R_{n_\nu}^{\nu}\notag
\end{align}
where $U_\pm^{(s)}(x,k_\nu):=\left. \cfrac{\partial^s}{\partial k^s } U_\pm(x,k,t)\right |_{k=k_\nu}$.
We call matrices $R_{n_\nu}^s(t)$ the norming matrices by an analogy with the norming constants in the scalar case.
	
The set  		
\begin{equation}\label{12}
 \mathcal S(t)=\{S_{12}(k,t),k_\nu,n_\nu,R_1^{\nu}(t),R_2^{\nu}(t),...,R_{n_\nu}^{\nu}(t), \;(\nu=1,2,...,N)   \}
\end{equation}
is called the scattering data for  equation (\ref{7}).
The aim of the direct scattering problem is to find these data for a given matrix $\Phi(x,t)$.
The inverse scattering problem consists in the reconstruction of the the matrix $\Phi(x,t)$  from  the scattering data of equation (\ref{7}).

The recovery procedure is the following.
Introduce the matrix-function
\begin{equation}\label{13}
  F(x,t)=\frac{1}{2 \pi}\int^\infty_{-\infty}S_{12}(k,t) e^{i k x}d k +\sum_{\nu=1}^N M_\nu(x,t) e^{i k_\nu x},
\end{equation}
where
\begin{equation*}
  M_\nu(x,t) = -\mathrm i \sum_{j=0}^{n_\nu - 1} \frac{(\mathrm i x)^j}{j !} R_{j+1}^{\nu}(t).
\end{equation*}
Consider the integral equation of the Fredholm type with respect to the variable $z\geq x$   for fixed $x$ and $t$:
\begin{equation}\label{14}
  A(x,z,t) - \int_x^\infty A(x,y,t) F(y+z,t) d y +\int_{x+z}^\infty F(y,t) d y =0.
\end{equation}
This equation has the unique matrix solution $A(x,z,t)$. The matrix $\Phi(x,t)$ is connected with this solution by formula
\begin{equation}\label{16}
  \Phi(x,t) = \cfrac{d A(x,x,t)}{dx} \left(\mathbb I - A(x,x,t)\right)^{-1}.
\end{equation}

Let  now $ \Phi(x,t)$ be the solution of equation (\ref{4}).
Then the scattering data evolve  with respect to $t$  as follows
\begin{equation}\label{scat}
  S_{12}(k,t) = S_{12}(k,0) e^{i 8 k^3 t}, \quad k_\nu(t)=k_\nu(0), \; n_\nu(t)=n_\nu(0),
\end{equation}
and the norming matrices $ R_l^{\nu}(t), \; l=1,2,...,n_\nu $  satisfy the differential equations
\begin{align}\label{15}
  &\frac{d}{d t} R_{n_\nu}^{\nu}(t) - 8 i k_\nu^3 R_{n_\nu}^{\nu}(t)  = 0, \notag \\
  &\frac{d}{d t} R_{n_\nu-1}^{\nu}(t) - 8 i k_\nu^3 R_{n_\nu-1}^{\nu}(t)  = 24 i k_\nu^2 R_{n_\nu}^{\nu}(t),\notag \\
  &\frac{d}{d t} R_{n_\nu-2}^{\nu}(t) - 8 i k_\nu^3 R_{n_\nu-2}^{\nu}(t)  = 24 i k_\nu^2 R_{n_\nu-1}^{\nu}(t) + 24 i k_\nu R_{n_\nu}^{\nu}(t),\\
  &.............................................................................\notag \\
  &\frac{d}{d t} R_{n_\nu-l}^{\nu}(t) - 8 i k_\nu^3 R_{n_\nu-l}^{\nu}(t)  = 24 i k_\nu^2 R_{n_\nu-l+1}^{\nu}(t) + 24 i k_\nu R_{n_\nu-l+2}^{\nu}(t) \notag \\
  &+ 8 i R_{n_\nu-l+3}^{\nu}(t), \quad
  l=3,4,...,n_\nu-1. \notag
\end{align}

This procedure allows us to solve the Cauchy problem for  equation (\ref{4}) with the initial matrix $\Phi_0(x)$.  Namely,
 let us solve the direct scattering problem for equation (\ref{7}) with the matrix $\Phi(x,0)=\Phi_0(x)$ and find the  scattering data $\mathcal S(0)$.
Then let us transform   these data with respect to $t$  according to (\ref{scat}), (\ref{15}), get $\mathcal S(t)$,  introduce the function \eqref{13},  solve the equation \eqref{14}, and obtain the solution  $\Phi(x,t)$  of the Cauchy problem by formula \eqref{16}.

In view of Lemma \ref{lem1} and the equivalence of conditions (\ref{3}) and (\ref{5}) the recovery procedure is valid also for the vector equation (\ref{2}).

We do not intend to study this Cauchy problem in details in this paper.
Our aim is to construct some exact solutions of equation (\ref{2}) starting from the scattering data.
To this end, we have to describe additional conditions on them such that the inverse scattering transform procedure would lead to a matrix (\ref{16}) satisfying (\ref{5}).
In a very general form  such additional conditions on the scattering data can be formulated as follows:
\begin{enumerate}[{\bf I.}]
\item {\it Poles $k_\nu$  of the transmission matrix $S_{11}(k,t)$  are located symmetrically with respect to  the imaginary axis and the norming matrices (\ref{12}) satisfy conditions
$R_p^{\nu}(t) = R_p^{\mu}( t)$, where $k_\nu=-\overline{k_\mu}$ and $p=1,...,n_\nu$ with $n_\nu=n_\mu$.}

\item {\it Matrix function $F(x,t)$  in (\ref{13}) can be represented as
\begin{equation}\label{17}
  F(x,t) = f_0(x,t)\mathbb I +\mathrm i[f_1(x,t)\sigma_1+f_2(x,t)\sigma_2+f_3(x,t)\sigma_3],
\end{equation}
where the functions  $f_l(x,t)$  are real and $\sigma_l$ are the Pauli matrices. }
\end{enumerate}
Further, put  $S_{12}(k,t)\equiv0$  and choose  $k_\nu, \; n_\nu$  and  $R_l^{\nu}$ to satisfy  {\bf I}, {\bf II} with $f_0(x,t)=0$.
Then the kernel of the integral equation (\ref{14}) is degenerated, and this equation is reduced to a linear algebraic system of equations which
can be solved exactly. Taking into account (\ref{17}) we get then the following representation for  $A(x,x,t)$
\begin{equation*}
  A(x,x,t)=a_0(x,t)\mathbb I+\mathrm i[a_1(x,t)\sigma_1+a_2(x,t)\sigma_2+a_3(x,t)\sigma_3],
\end{equation*}
where $a_l(x,t), \; l=0,1,...,3$ are real-valued functions, which can be found explicitly and satisfy equality
\[\sum_{l=0}^3 a_l^2(x,t)= 2a_0(x,t). \]
Then by \eqref{3} and \eqref{vec} the solution of  $\mathbf u(x,t)$ of (\ref{2}) can be determined via the formulas
\begin{align*}
  \mathbf u_1= a_2^\prime a_3 - a_3^\prime a_2 - a_1^\prime (1-a_0) - a_1 a_0^\prime, \\
  \mathbf u_2= a_3^\prime a_1 - a_1^\prime a_3 - a_2^\prime (1-a_0) - a_2 a_0^\prime,   \\
  \mathbf u_3= a_1^\prime a_2 - a_2^\prime a_1 - a_3^\prime (1-a_0) - a_3 a_0^\prime,
\end{align*}
with $ a_l^\prime :=  \cfrac{\partial a_l}{\partial x} (x,t), \quad l=1,2,3 $.

In the next section we apply this approach to construct  exact solutions of the vector modified Korteweg-de Vries equation (\ref{2}).

\section{Analytical solutions of the vector mKdV equation.}
\subsection{Single soliton solution. }
We begin with the simplest case.
Assume that the transmission matrix $S_{11}(k,t)$   has a simple pole at the point $k_1=i\mu, (\mu>0, n_1=1)$.
Choose the matrix $R_1^{1}(0)$  in the form
\begin{equation}\label{18}
  R_1^{1}(0)=\alpha\sigma_1
  +\beta\sigma_2+\gamma\sigma_3	= \begin{pmatrix} \gamma & \alpha-\mathrm i \beta \\ \alpha+\mathrm i \beta & -\gamma \end{pmatrix}	
\end{equation}
where  $\alpha, \beta, \gamma$  are real constants.
Then the conditions {\bf I} and {\bf II} are satisfied.
Solving equation (\ref{14}) with  the kernel $F(x,t)=-\mathrm i e^{-\mu x} R_1^{1}(t)$  we get the following solution
\begin{equation*}
  A(x,y,t)=\mathrm i\frac{R_1^{1}(0)}{\mu} e^{-\mu (x+y-8 \mu^2)t} \left(1+\mathrm i\frac{R_1^{1}(0)}{2 \mu} e^{-\mu (2 x-8 \mu^2)t}\right)^{-1}.
\end{equation*}
Denote $\mathbf {v}=(\alpha, \beta, \gamma)$, $\|\mathbf{v}\|=\sqrt{\alpha^2 + \beta^2+\gamma^2}$. Taking into account (\ref{16}), we represent the solution of the matrix equation  (\ref{4}) as
\begin{equation}\label{phi}
  \Phi(x,t)=-\mathrm i\frac{2\mu}{\ch\left(2\mu(x-4\mu^2t)-\ln\|\mathbf{v}\|+\ln 2\mu\right)}  \frac{R_1^{1}(0)}{\|\mathbf{v}\|}.
\end{equation}
Denote now $ a=2\mu$, $c=a^2$, $\varphi=\cfrac{1}{a} \ln\left(\cfrac{a}{\|\mathbf{v}\|}\right)$,  $\mathbf {e}=\cfrac{\mathbf v}{\|\mathbf{v}\|}$.

According to (\ref{3}), \eqref{vec},  (\ref{18}) and (\ref{phi}), the vector function
\begin{equation}\label{19}
  \mathbf u(x,t)=\frac{a}{\ch(a(x-ct)-\varphi)}\mathbf e
\end{equation}
is a single-soliton solution of the vector equation mKdV (\ref{2}).
This soliton has an amplitude  $a$, speed  $c$, phase $\varphi$  and is directed along the vector  $\mathbf e$.
This direction does not depend on the spatial and time variables.
Note that the scalar function in front of the vector $\mathbf e$ in \eqref{19}  is the well known  single-soliton solution to the scalar equation mKdV  (see \cite{2}).

\subsection{Solutions of the soliton-antisoliton type.}
Consider now the case of double pole  $k_1=i\mu$, $\mu>0,$ $ n_1=2$.
Choose the matrices  $R_1^{1}(0)$, $R_2^{1}(0)$   as
\begin{equation*}
    R_1^{1}(0) = \alpha_1\sigma_1+\beta_1\sigma_2+\gamma_1\sigma_3,\quad
    R_2^{1}(0) = \mathbf i(\alpha_2\sigma_1+\beta_2\sigma_2+\gamma_2\sigma_3),
\end{equation*}
where $\alpha_l, \beta_l, \gamma_l $,  $l=1,2$ are real constants. Denote
\begin{equation}\label{vi} \mathbf v_l=(\alpha_l, \beta_l, \gamma_l), \quad l=1,2.\end{equation}
Since the kernel of the integral equation (\ref{14}) has the form
\begin{equation*}
  F(x,t)=-\mathrm i(R_1^{1}(t)+\mathrm ixR_2^{1}(t)) e^{-\mu x},
\end{equation*}
then the conditions {\bf I}, {\bf II} are satisfied.
 In this case the described above approach  leads to a solution  $\mathbf {u}(x,t)$ to  (\ref{2}) of the type soliton-antisoliton. This solution is sometimes called doublet.

Consider first the special case   $R_1^{1}(0)=0$. Put $a=2\mu$, $c= a^2$ and $\phi=\ln\dfrac{c}{\|\mathbf v_2\|}$.
After the change  of  variables
\begin{equation*}
  x\rightarrow x+\frac{1+3\varphi}{2 a},\; t\rightarrow t+\frac{1+\varphi}{3 a c}
\end{equation*}
this doublet can be represented as
\begin{equation}\label{21}
  \mathbf u(x,t)=2 a \frac{a \sh(a(x-c t))(3 c t - x)+\ch(a(x-c t))}{\ch^2(a(x-c t))+a^2(3 c t - x)^2} \mathbf{e}_2,
\end{equation}
where $\mathbf e_2=\dfrac{\mathbf v_2}{\|\mathbf v_2\|}$ (cf. \eqref{vi}). 	

\begin{figure}[ht!]
  \begin{center}
     \includegraphics[width=0.8\linewidth, keepaspectratio]{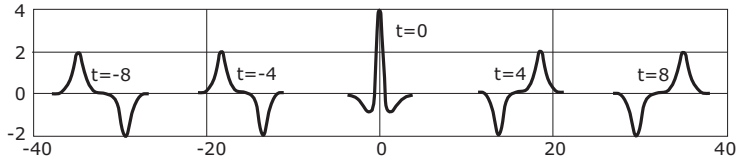}
     \caption{The motion phases of the doublet  ($R_1^{1}(0)=0$,\; $\mu=1$)}
    \label{r1}
  \end{center}
\end{figure}

For sufficiently large time the doublet splits up into a sum of oppositely directed solitons with the amplitudes  $a$ (the soliton and the anti-soliton).
The distance between them grows logarithmically with respect to time
\begin{equation*}
 \mathbf u(x,t)= \mathbf e_2 \left\{\begin{aligned}
  & \frac{a}{\ch(a(x-c t) - \ln(4 a^3 t))} - \frac{a}{\ch(a[x-c t] + \ln(4 a^3 t))},&t\rightarrow\infty \\
  & -\frac{a}{\ch(a(x-c t) - \ln(4 a^3 |t|))} + \frac{a}{\ch(a(x-c t) + \ln(4 a^3 |t|))},&t\rightarrow -\infty
  \end{aligned}\right.
\end{equation*}
When the time increases  the  soliton catches up and then overtake the anti-soliton.
At their closest approach and  interaction there occur the  resulting impulse
\begin{equation*}
  \mathbf u(x,0)=2 a \frac{\ch(a x)-a x \sh(a x)}{\ch^2(a x) + (a x)^2} \mathbf e_2.
\end{equation*}
The motion phases of the doublet are show on  fig. \ref{r1}.

The constructed vector solution \eqref{21} is  in  one-component, because  it is directed for any $x\in\mathbb R$  and $t\in\mathbb R$  along the constant vector  $\mathbf e_2$.
The scalar function in front of  $\mathbf e_2$ in \eqref{21}  satisfies scalar mKdV equation \eqref{1}, for which a  doublet type solution is well known (\cite{3}, \cite{7}).

Consider now   more general case. Suppose that  $R_1^{1}(0)\neq 0 $  and  this matrix is not proportional to $R_2^{1}(0)$. Then the doublet type solution  loses the property to be one-component and  the vector equation (\ref{2}) is not reduces  to the scalar equation mKdV. Indeed, suppose that the unit vectors
 $\mathbf e_1 = \dfrac{\mathbf v_1}{\|\mathbf v_1\|}$  and    $\mathbf e_2 = \dfrac{\mathbf v_2}{\|\mathbf v_2\|}$ are  orthogonal.  Denote
 \begin{equation}\label{del}\delta=\frac{\|\mathbf v_1\|}{\|\mathbf v_2\|}, \quad \rho = \cfrac{a \delta}{2} .\end{equation} Then  in the shifted coordinate system
\begin{equation*}
    x\rightarrow x+\frac{1+3\varphi}{2 a} - \frac{\delta}{2},t\rightarrow t+\frac{1+\varphi}{3 a c} - \frac{\delta}{4 c},
\end{equation*}
this doublet has the form
\begin{equation}\label{dupl}
   \mathbf u(x,t)=u_1(x,t)\mathbf e_1 + u_2(x,t) \mathbf  e_2 + u_3(x,t) (\mathbf e_1 \times \mathbf e_2),
\end{equation}
where
\begin{multline*}
    u_1(x,t)= 2 a \rho
         \bigg\{\frac{-\ch(a(x-c t))}{\ch^2(a(x-ct))+a^2(3 c t - x)^2 +\rho^2} + \\ +4 \frac{\ch(a(x-ct))-a (3 c t - x) \sh(a(x- ct))}{\left( \ch^2(a(x- ct))+a^2(3 c t - x)^2 +\rho^2      \right)^2}  \bigg  \}
\end{multline*}
\begin{multline*}
    u_2(x,t)= 2 a \bigg \{ \frac{\ch(a(x- c t))+a (3 c t - x) \sh(a(x- c t))}{\ch^2(a(x- ct))+a^2(3 c t - x)^2 + \rho^2} - \\ -
          \frac{\rho^2  \ch(a(x-ct))}{\left( \ch^2(a(x-ct))+a^2(3 c t - x)^2 +\rho^2     \right)^2}\bigg \}
\end{multline*}
\begin{multline*}
    u_3(x,t) = -4 a \rho \frac{2 a (3 c t - x)+\ch(2 a (x- c t)}{{\left(\ch^2(a(x- ct))+a^2(3 c t - x)^2 +\rho^2      \right)^2}}.\\
\end{multline*}

\begin{figure}[!ht]
  \begin{center}
    \includegraphics[width=0.8\linewidth, keepaspectratio]{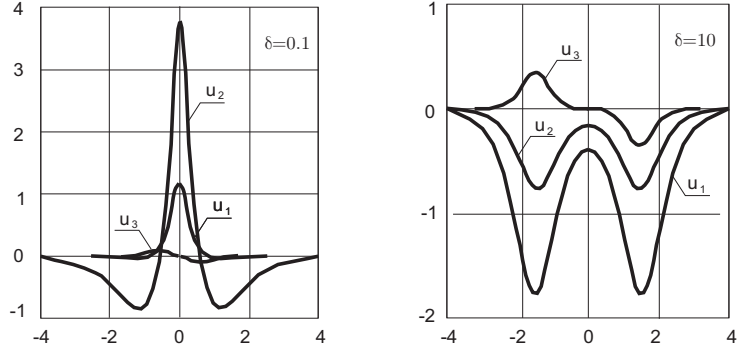}
    \caption{The components of the resulting impulse of the doublet  ($\mu=1$) at $\delta=0.1$  and $\delta=10$.}
    \label{r2}
  \end{center}
\end{figure}

Asymptotic analysis of these formulas as  $t\rightarrow\pm\infty$ shows that doublet \eqref{dupl} splits up onto two oppositely directed solitons with the amplitude $a$ oriented along the vector  $\mathbf e_2$. These solitons draw  together as $|t|\to 0$ and interact with each other. For \ $t=0$ they produce the resulting three-component impulse depending on $\delta$  (cf. \eqref{del}), as is shown on  fig. \ref{r2}.

	Similarly, one can consider the case of one pole  $k_1=i\mu, (\mu>0)$ with  higher multiplicity  $n_1\geq3$. Analytical formulas in this case are quite combersome. That is why for the case  $n_1=3$  we perform numerical computations based on  the inverse scattering transform given by  (\ref{12}) - (\ref{16}). We choose  norming matrices as $R_l^{1}(0)=(\mathrm i)^{l-1}(\alpha_l\sigma_1+\beta_l\sigma_2+\gamma_l\sigma_3)$, where $\alpha_l,\beta_l,\gamma_l\in \mathbb R$, $l=1,2,3$. One can check that in this case the conditions $\mathbf I$, $\mathbf {II}$ are also satisfied.
	
	Numerical computations show that in this case the solution of \eqref{2} is of the type  soliton-antisoliton-soliton,  and one can call it the triplet. For large times the triplet split up onto  three mutually oppositely directed solitons of the same amplitude. The distance between them grows logarithmically with respect to time as $t\rightarrow\pm\infty$.
When $|t|\to 0$ these solitons merge into the  resulting impulse. In the general case when the vectors $\mathbf{v}_l=(\alpha_l,\beta_l,\gamma_l  ),\; l=1,2,3$  are not collinear, this impulse  has three components  depending on  $x$  and  $t$.  If the vectors $\mathbf v_l$   are all collinear, then the triplet  is one-component. This case is illustrated by fig. \ref{r3}:
\begin{figure}[ht!]
  \begin{center}
    \includegraphics[width=0.8\linewidth, keepaspectratio]{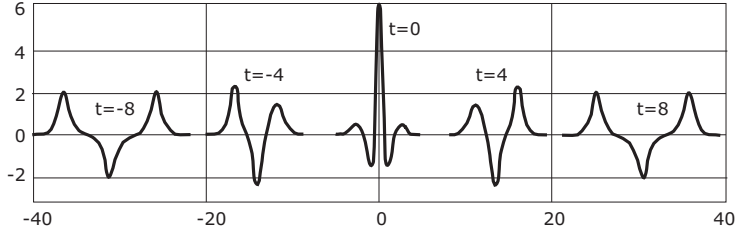}
    \caption{The motion phases  of the triplet   ($R_1^{(1)}(0)=R_2^{(1)}(0)=0$,\;$\mu=1$).}
    \label{r3}
  \end{center}
\end{figure}

\subsection{The breather type solutions. }
\begin{figure}[ht!]
  \begin{center}
    \includegraphics[width=0.8\linewidth, keepaspectratio]{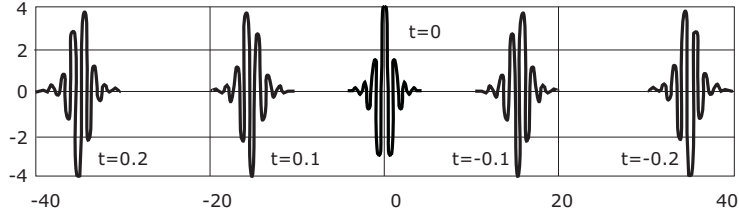}
 \caption{The motion phases of the breather for $\mu=1 , \; \lambda=4$.}
    \label{r4}
  \end{center}
\end{figure}

Consider now the case of two simple poles $k_{1,2}=\pm\lambda+i\mu, (\lambda,\mu>0)$  and let  $R_1^{(2)}=R_2^{(2)}=\alpha\sigma_1+\beta\sigma_2+\gamma\sigma_3$, where $\alpha,\beta,\gamma\in\mathbb R$. Then  $$F(x,0)=-2i(\alpha\sigma_1+\beta\sigma_2+\gamma\sigma_3) e^{-\mu x}cos(\lambda x),$$  and one can check that the conditions {\bf I}, {\bf II} are satisfied. In this case we obtain the multisoliton solution of \eqref{3} of the breather type.
Put $\mathbf v=(\alpha, \beta, \gamma)$, \[c_\mu=4(\mu^2-3 \lambda^2),\ c_\lambda=4(3 \mu^2-\lambda^2), \ \psi=\arctan \cfrac{\lambda}{\mu},\
\varphi = \ln\left(\cfrac{4\mu^2}{\|\mathbf v\| } \cfrac{|\mu^2-\lambda^2|}{\mu^2+\lambda^2}\right).\] Then in the shifted coordinate system
\begin{equation*}
    x\rightarrow x +\frac{1}{2}\left(\frac{\psi}{\lambda}c_\mu - \frac{\varphi}{\mu}c_\lambda\right)\frac{1}{c_\mu-c_\lambda},
t\rightarrow t +\frac{1}{2}\left(\frac{\psi}{\lambda} - \frac{\varphi}{\mu}\right)\frac{1}{c_\mu-c_\lambda},
\end{equation*}
the breather type solution can be represented as  (see fig. \ref{r4})
\begin{equation*}
 \mathbf u(x,t)= \frac{\lambda \cos(2\lambda(x - c_\lambda t)) \ch(2 \mu (x-c_\mu t))
    +\mu \sin(2\lambda(x - c_\lambda t)) \sh(2 \mu (x-c_\mu t))}{\dfrac{\lambda}{4\mu}\ch^2(2\mu(x - c_\mu t))+\dfrac{\mu}{4\lambda}\sin^2(2\lambda (x - c_\lambda t))} \mathbf e,
\end{equation*}
where $\mathbf e = \dfrac{\mathbf v}{\|\mathbf v\|}$.

When  $\lambda\rightarrow0$  this solution degenerates into a doublet, and when $\mu\ll\lambda$  one gets the soliton with  an internal motion
\begin{equation}\label{int}
    \mathbf u(x,t)=4\mu\frac{\sin(2\lambda(x-c_\lambda t))}{\ch(2\mu(x-c_\mu t))} \mathbf{e}, \quad c_\mu=3 c_\lambda<0.
\end{equation}
	
\begin{figure}[!ht]
  \begin{center}
    \includegraphics[width=0.8\linewidth, keepaspectratio]{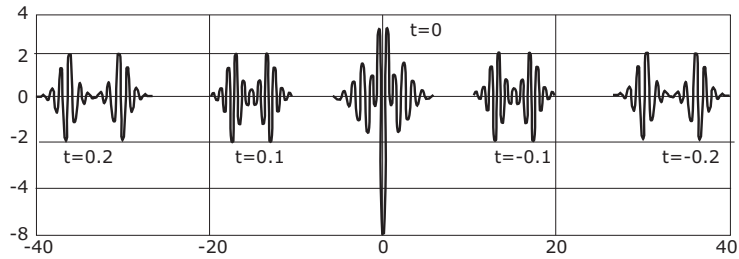}
    \caption{The motion phases of the  linked breathers    with $\mu=1 , \; \lambda=4$.}
    \label{r5}
  \end{center}
 \end{figure}

	This solution is one-component, since   $\mathbf e$  is a constant vector. Therefore  solution \eqref{int} coincides with a solution of the breathers type for the scalar mKdV equation.
	However, if the poles $k_{1,2}=\pm\lambda+i\mu$ are multiple, the solution loses its property to be one-component, and the vector equation (\ref{2}) is not reduct to a scalar equation. Analytical formulas for the solution in this case are quite cumbersome. That is why we compute the solution of \eqref{2}  numerically. Namely, assume that the pole $k_1=\lambda + \mathrm i \mu$ has multiplicity 2, and respective two norming matrices are represented as
  $R_l^{1}(0)=\alpha_l\sigma_1+\beta_l\sigma_2+\gamma_l\sigma_3$,
where $\alpha_l,\beta_l,\gamma_l\in \mathbb R$, $l=1,2$.  According the condition {\bf I}, the second pole $k_2=-\lambda + \mathrm i \mu$ is  double also, with the same norming matrices. The condition {\bf II} is also fufiled due to the structure of these matrices.  The simulation shows that for $t\rightarrow \pm\infty$ the solution splits  into two one-component   breathers, and the distance between them increases logarithmically with time. When they approach the resulting impulse has three-component character
 in the case when the vectors $ \mathbf v_l=(\alpha_l, \beta_l, \gamma_l)$, $l=1,2$   are not collinear. If the vectors  $ \mathbf v_l$ are collinear, then the vector solution is one-component and   it is directed for all $x,t$ along the same vector. The  graph of the scalar function in front of this vector is show on fig. \ref{r5}.

\subsection{Interacting solitons.}
Consider now the case of two simple pure imaginary poles of the transmission matrix $S_{11}(k,0)$: $k_j=\mathrm i\mu_j, $\;\ $\mu_1>\mu_2>0$. As usual we choose the norming matrices as  $R_1^{l}(0)=\alpha_l\sigma_1+\beta_l\sigma_2+\gamma_l\sigma_3, \;l=1,2$ , where  $\alpha_l,\beta_l,\gamma_l\in\mathbb R$. It is clear that the condition {\bf I}, {\bf II} are fulfilled.  Assume first that the vectors $\mathbf v_l=(\alpha_l, \beta_l,\gamma_l)$  are collinear , that is \begin{equation}\label{DE}\mathbf e:=\frac{\mathbf v_1}{\|\mathbf v_1\|}=\varepsilon\frac{\mathbf v_2}{\|\mathbf v_2\|}, \quad \varepsilon=\pm 1.\end{equation}  Put
\begin{equation}\label{not}a_j=2\mu_j,\ c_j=a_j^2,\ \varphi_j=\ln\left( \frac{a_j}{\|\mathbf v_j\|} \right), \ y_j := a_j (x - c_j t), \quad j=1,2.\end{equation} Then the change of variables \begin{equation}\label{211}
       x\rightarrow x - \frac{1}{c_1-c_2}\left(\frac{\varphi_2}{a_2}c_1 - \frac{\varphi_1}{a_1}c_2   \right),
     t\rightarrow t + \frac{1}{c_1-c_2}\left(\frac{\varphi_2}{a_2} - \frac{\varphi_1}{a_1}   \right),
  \end{equation}
leads to the following solution of \eqref{2}:
\begin{equation}\label{22}
         \mathbf u(x,t)=\frac{2 (a_1^2 - a_2^2) (a_1 \ch(y_2) + \varepsilon a_2 \ch(y_1))}{4 \varepsilon a_1 a_2 +(a_1+a_2)^2 \ch(y_2-y_1) + (a_1-a_2)^2 \ch(y_2+y_1)}\mathbf e.
       \end{equation}

 One can see that for large  $|t|$  this solution splits into two non-interacting solitons
\begin{equation}\label{psik}
    \mathbf u(x,t)= \mathbf  e \left\{\begin{aligned}
& \frac{a_1}{\ch\left(y_1  - \psi \right)} + \frac{\varepsilon a_2}{\ch\left(y_2  + \psi \right)},
 \; &t\rightarrow\ -\infty, \\
& \frac{a_1}{\ch\left(y_1  + \psi \right)} + \frac{\varepsilon a_2}{\ch\left(y_2  - \psi \right)},
 \; &t\rightarrow\infty,
\end{aligned}\right.
\end{equation}
where $\psi =  \ln\dfrac{a_1+a_2}{a_1-a_2},\quad a_1>a_2$.	

\begin{figure}[!ht]
  \begin{center}
    \includegraphics[width=0.8\linewidth, keepaspectratio]{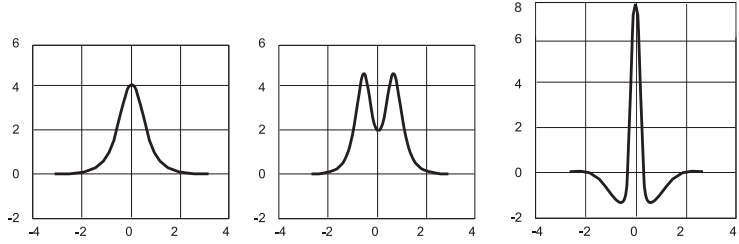}
     \caption{The resulting impulse of the collinear solitons: the merge-split type ($\mu_1=3,\mu_2=1$) and the bounce-exchange type ($\mu_1=3,\mu_2=2$) for unipolar solitons and the absorb-emit type ($\mu_1=3,\mu_2=1$) for heteropolar ones.}
    \label{r6}
  \end{center}
\end{figure}

When these solitons approach, they begin to interact strongly  with the resulting impulse\begin{equation}\label{*}
      \mathbf u(x,0)=\frac{2 (a_1^2 - a_2^2) (a_1 \ch(a_2 x) + \varepsilon a_2 \ch(a_1 x))}{4 \varepsilon a_1 a_2 +(a_1+a_2)^2 \ch(a_2 x - a_1 x) + (a_1-a_2)^2 \ch(a_2 x + a_1 x)}\mathbf e,
 \end{equation}
whose graph is shown in fig. \ref{r6}.

Since the solution \eqref{22}  is one-component and is directed along
the vector $\mathbf  e$ independent  on $x,t$ then a solution of scalar mKdV equation \eqref{1} corresponds to it. This scalar solution is constructed in \cite {3}. Note, that in \cite{3}  an interesting interpretation of the resulting impulse graph is given. The graph is treated  as a characteristics of the interaction type for solitons. Namely, the resulting impulse with two humps corresponds to the bounce-exchange interaction of the unipolar solitons. This type of interaction occurs when the amplitudes do not differ much: $\dfrac{ 3-\sqrt{5}}{2} a_1< a_2< a_1$. In this case, the solitons do not stick, but the faster soliton gives its power to the  slower soliton, then  the faster soliton is decelerated, and the other one  is accelerated. The impulse with one hump corresponds to the merge-split type of the unipolar solitons interaction with a larger difference of the amplitudes  $a_2<\dfrac {3-\sqrt{5}}{2}  a_1$. In this case, the faster soliton absorbs the slower solitons during interaction and then restores. The impulse with three  extrema corresponds to the interaction of the heteropolar solitons, when the so called absorbance-emit mode is  realized (see \cite {3}).

	Consider the case when the vectors  $\mathbf v_l=(\alpha_l,\beta_l,\gamma_l )$, $l=1,2$ are  not collinear, moreover, they are orthogonal. Then the solution of the vector equation mKdV \eqref{2} is essentially  the three component.  Namely,  denote $\mathbf e_l=\dfrac{\mathbf v_l}{\|\mathbf v_l\|}$. Then in the shifted coordinate system \eqref{211} the solution of \eqref{2} is given by formula
\begin{equation*}
     \mathbf u(x,t) =  u_1(x,t)  \mathbf e_1 +  u_2(x,t)  \mathbf e_2 +   u_3(x,t) (\mathbf e_1\times  \mathbf e_2),
\end{equation*}
where
\begin{multline*}
 u_1(x,t) = (a_1^2-a_2^2) a_1 \\
  \times  \frac{(a_1 - a_2)^2 \ch(2 y_2 + y_1) + (a_1 + a_2)^2 \ch(2 y_2-y_1) + 2 (5 a_2^2 - 3 a_1^2) \ch(y_1 ) }{ ( (a_1 - a_2)^2 \ch( y_2 +y_1 ) + (a_1 + a_2)^2 \ch( y_2 -y_1)  )^2 },
\end{multline*}

\begin{multline*}
u_2(x,t) = (a_1^2-a_2^2) a_2 \\
 \times   \frac{(a_1 - a_2)^2 \ch( y_2 + 2 y_1) + (a_1 + a_2)^2 \ch( y_2 - 2 y_1) + 2 (5 a_1^2 - 3 a_2^2) \ch(y_2 ) }{ ( (a_1 - a_2)^2 \ch( y_2 +y_1 ) + (a_1 + a_2)^2 \ch( y_2 -y_1)  )^2 },
\end{multline*}

\begin{multline*}
u_3(x,t) = 4 (a_1^2-a_2^2)^2 \frac{ (a_1 - a_2) \sh(y_1  + y_2 ) +(a_1 + a_2) \sh(y_1  - y_2 ) }{ ( (a_1 - a_2)^2 \ch( y_2 +y_1 ) + (a_1 + a_2)^2 \ch( y_2 -y_1)  )^2 }.\\
\end{multline*}

As $|t|\rightarrow\infty$ this solution splits onto two noninteracting solitons, oriented in the orthogonal directions $\mathbf e_1$ and $\mathbf e_2$

\begin{equation} \label{sol5}
    \mathbf u(x,t)=  \left\{\begin{aligned}
&\mathbf e_1 \frac{a_1}{\ch\left(y_1  - \psi \right)} + \mathbf e_2 \frac{a_2}{\ch\left(y_2  + \psi \right)},
 \; &t\rightarrow\ -\infty ,\\
&\mathbf e_1 \frac{a_1}{\ch\left(y_1  + \psi \right)} + \mathbf e_2 \frac{a_2}{\ch\left(y_2  - \psi \right)},
 \; &t\rightarrow\infty,
\end{aligned}\right.
\end{equation}
where $ y_j$ and $\psi$ are defined by \eqref{not} and \eqref{psik} respectively.

\begin{figure}[ht!]
  \begin{center}
    \includegraphics[width=0.8\linewidth, keepaspectratio]{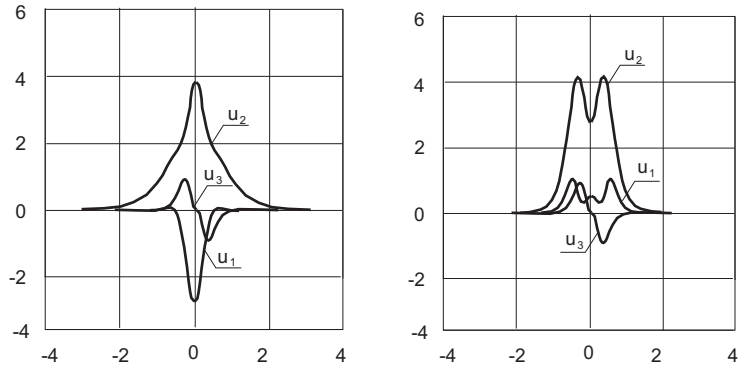}
   \caption{The components of the resulting impulse ($\mu_1=3,\mu_2=1$ and $\mu_1=3,\mu_2=2$).}
    \label{r7}
  \end{center}
\end{figure}

When $|t|\rightarrow0$ these solitons  interact and form three-component resulting impulse, as is show on fig. \ref{r7}.

\begin{figure}[ht!]
\begin{center}
     \includegraphics[width=0.8\linewidth, keepaspectratio]{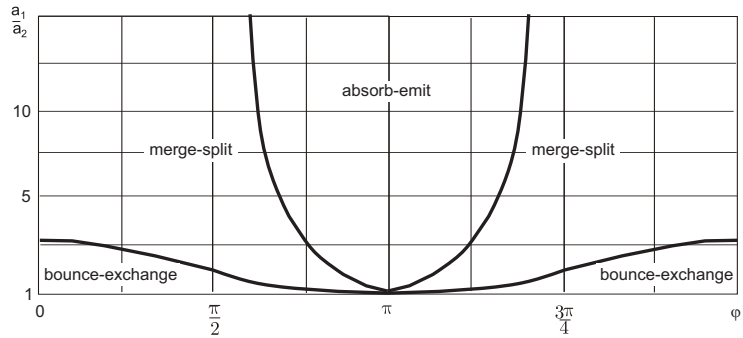}
 \caption{Types of two-soliton interactions depending on the amplitudes and directions
 ($\varphi$ is angle between vectors $\mathbf e_1$ and $\mathbf e_2$).}
    \label{r8}
      \end{center}
\end{figure}

If the vectors  $\mathbf v_j$  are not orthogonal, the analytical formulas are quite complicated. The analysis of the resulting three component impulse was given numerically. We found that the module of its  graph  can   have one, two or three maximums, depending on the ratio of the amplitudes of the solitons and their directions. Folloving \cite{3} it is naturally to call this cases as merge-split, bounce-exchange and absorbance-emit interactions (fig. \ref{r8}).

The examples considered above  demonstrate that one can build up different types of solutions for the vector mKdV by choosing singularities of the transmission matrix $S_{11}(k,0)$. If we choose the transmission matrix with the simple pure imaginary poles then we get a multisoliton solution. If the poles are multiple then we get the solution of the duplet types, triplet ets, i.e. witch are connected solitons with the growing with respect to time distance between them. The choice of the transmission matrix with the simple complex pole symmetric with respect to the imaginary axis leads to the solutions of the breather type, i.e. to the solitons with the inner motion. If the poles are multiple then we get a solution of the type of the connected breathers. The distance between them grows logarithmically with respect to time.

Moreover if the norming matrices $R_p^{\nu}(0)$ are generated by the non collinear vectors then all such solutions are essentially three-component.

\end{document}